\documentclass[sigconf]{acmart}

\AtBeginDocument{%
  }

\copyrightyear{2025}
\acmYear{2025}
\acmDOI{XXXXXXX.XXXXXXX}

\acmBooktitle{Under review} 
\acmISBN{978-1-4503-XXXX-X/18/06}

\usepackage{subcaption}
\usepackage{multirow}
\begin{document}

\title[FedFlex]{FedFlex: Federated Learning for Diverse Netflix Recommendations}


\author{Sven Lankester}
\authornotemark[1]
\affiliation{%
  \institution{Vrije Universiteit Amsterdam}
  \country{Netherlands}}
\email{s.lankester@student.vu.nl}

\author{Gustavo de Carvalho Bertoli}
\affiliation{%
  \institution{Airbus Central R\&T}
  \country{Germany}}
\email{gustavo.bertoli@airbus.com}

\author{Matias Vizcaino}
\affiliation{%
 \institution{Georgia Institute of Technology}
 \country{Mexico}}
\email{matiasvc8@gmail.com}

\author{Emmanuelle Beauxis Aussalet}
\affiliation{%
  \institution{Vrije Universiteit Amsterdam}
  \country{Netherlands}}
\email{e.m.a.l.beauxisaussalet@vu.nl}

\author{Manel Slokom}
\authornotemark[1]
\affiliation{%
  \institution{Centrum Wiskunde \& Informatica}
  \country{Netherlands}}
\email{manel.slokom@cwi.nl}
\authornote{*: The corresponding authors.}


\renewcommand{\shortauthors}{Lankester et al.}

\begin{abstract}


    The drive for personalization in recommender systems creates a tension between user privacy and the risk of ``filter bubbles''. Although federated learning offers a promising paradigm for privacy-preserving recommendations, its impact on diversity remains unclear. 
    We introduce FedFlex\footnote{SVD: \url{https://anonymous.4open.science/r/syftbox-netflix-svd-6661}}\footnote{BPR: \url{https://anonymous.4open.science/r/syftbox-netflix-bpr-CB5B}}, a two-stage framework that combines local, on-device fine-tuning of matrix factorization models (SVD and BPR) with a lightweight Maximal Marginal Relevance (MMR) re-ranking step to promote diversity. 
    We conducted the first live user study of a federated recommender, collecting behavioral data and feedback during a two-week online deployment.
    Our results show that FedFlex successfully engages users, with BPR outperforming SVD in click-through rate. Re-ranking with MMR consistently improved ranking quality (nDCG) across both models, with statistically significant gains, particularly for BPR. Diversity effects varied: MMR increased coverage for both models and improved intra-list diversity for BPR, but slightly reduced it for SVD, suggesting different interactions between personalization and diversification across models. 
    Our exit questionnaire responses indicated that most users expressed no clear preference between re-ranked and unprocessed lists, implying that increased diversity did not substantially reduce user satisfaction.

\end{abstract}

%
%


\keywords{Federated Learning, Recommender Systems, Diversity, Privacy, SyftBox, Live User Study}

\maketitle

\section{Introduction}
Top-N recommender systems (RSs) influence much of how we experience the online world and aim to solve the information overload problem. By training machine learning models on large amounts of user data, it becomes possible to turn the database of all possible Netflix series into an ordered list of what a user is most likely to watch. However, highly personalized RSs (called over-personalization) can have negative side effects. Firstly, processing a large amount of user data comes with privacy concerns~\cite{slokom2019data,slokom2021PerBlur}. In the case of Netflix, the user is not in control of what parts of their viewing history they share with the company. Secondly, by creating highly personalized recommendations a \textit{filter bubble} can be created. This is an effect in which users only receive recommendations for what they already like, which can stifle exploration and reinforce existing preferences and biases~\cite{olddiversity}.

Federated learning is a machine learning paradigm that preserves privacy by training models on user data locally and sharing only the updated parameters with a global model~\cite{Qiang2019Federated,FedRankLowTraining}. Recommendations can then be generated from the global model without ever sharing user data~\cite{FedRec,usercontrolfedtopn}. Privacy can be further enhanced using techniques like \textit{differential privacy}, which adds noise to model updates to prevent inference of user data from the shared parameters~\cite{Dwork2008Differential}.
Several studies explore different methods for federated recommendation. One focuses on reducing communication costs through a framework that lowers overhead when transmitting model updates~\cite{FedRankLowTraining}, while another emphasizes privacy, allowing users to control the amount of sensitive data shared~\cite{usercontrolfedtopn}. However, few works address objectives beyond accuracy, such as fairness and diversity. For fairness, \textit{FairFed} promotes group fairness via a fairness-aware aggregation method~\cite{FairFed}, and for diversity, \textit{FDRS}~\cite{FDRS} uses a graph-based approach to promote diversity in a federated setting.

Our study addresses a gap in the literature by conducting a user study with live participants on a federated learning-based recommender system, focusing on promoting diversity. The main contribution lies in our approach, implementation, and evaluation of a federated recommender in a live setting. We investigate the following research questions (RQs):  
\textbf{RQ1:} How can a reproducible federated recommender be designed?  
\textbf{RQ2:} To what extent can diversity be promoted in a federated setup?  
\textbf{RQ3:} How can a diverse federated recommender be evaluated in a live user study?

\section{Background and Related Work}

In this section, we briefly cover works related to recommendation based on federated learning.

\subsection{Federated Recommendations}

Federated recommendation has emerged as a privacy-focused alternative to traditional centralized recommender systems. Early signs of the potential of federated learning can be traced back to a blog post from researchers at Google~\cite{GoogleBlogPost} testing out federated learning technologies and discussing progress they have made in developing the underlying algorithms. In the literature, existing recommender technologies are commonly adapted for a federated environment. For example, Federated Collaborative Filtering (FCF)~\cite{FCF} proposes a framework with the goal of ranking items with implicit feedback. This is achieved by sharing gradient updates instead of raw data, obscuring the user data on which the model is trained. FedRec~\cite{FedRec} follows up on FCF by proposing a generic federated learning framework for rating prediction with explicit feedback. Their framework is flexible and can be applied to a wide variety of factorization-based models.

\subsection{Diversity in Federated Learning}

Research on diversity within federated learning is sparse. Existing works often only address one aspect of fairness, and there is only one work focusing on diversity, to the best of our knowledge. One such work is the aforementioned FairFed~\cite{FairFed}. FairFed aims to tackle the problem of group fairness, a problem often solved by having access to sensitive information, which is not the case in federated learning. In their approach, they train a model using fairness-aware aggregation with local debiasing and then evaluate the fairness of the global model on a local dataset. Their work raises the concern of a difference between local and global fairness due to the distributed nature of the system.
FDRS~\cite{FDRS}, on the other hand, focuses on introducing diversity into federated learning. Their proposed framework for a Federated Diversified Recommender System (FDRS) is based on heterogeneous graph convolutional networks (HGCN). They first construct heterogeneous user-item graphs, based on which the HGCN learns rich user and item representations. Clients can then generate diversity-aware recommendations locally.

This section details our methodology for building and evaluating FedFlex. We begin by describing its architecture, which is based on SyftBox\footnote{SyftBox: \url{https://github.com/OpenMined/syftbox}}. Next, we describe the core recommendation algorithms, including our approach to promoting diversity, and finish with the design of the live user study.

\subsection{SyftBox}

Our methodology builds on SyftBox, an open-source protocol based on PySyft~\cite{pysyft} for privacy-preserving computations on distributed datasets. 
SyftBox enables federated learning without centralizing sensitive user data.
We selected SyftBox to operationalize our recommender system because it directly addresses our core challenge: training on interaction data (e.g., viewing history, clicks) while keeping the raw data on users' devices. SyftBox coordinates secure communication between the central aggregator and participants, exchanging only model updates rather than private data.
SyftBox organizes user data into \textit{datasites} (network directories) where users retain fine-grained permission control. While it does not natively implement Privacy-Enhancing Technologies (PETs) such as differential privacy or homomorphic encryption, it provides the infrastructure to apply them. Applications are open source, allowing users to verify trustworthiness before granting access.
Alternatives like Flower\footnote{https://flower.ai/} and syft\_flwr\footnote{https://github.com/OpenMined/syft-flwr} exist; we chose SyftBox due to prior experience and its ease of deployment. For our live study, SyftBox lowered the entry barrier by abstracting federated learning complexities: participants simply installed a packaged application, while SyftBox handled setup and synchronization~(\textbf{RQ1}).

\subsection{FedFlex}

\label{Sec:App}

\begin{figure}[!h]
\centering
  \includegraphics[width=\columnwidth]{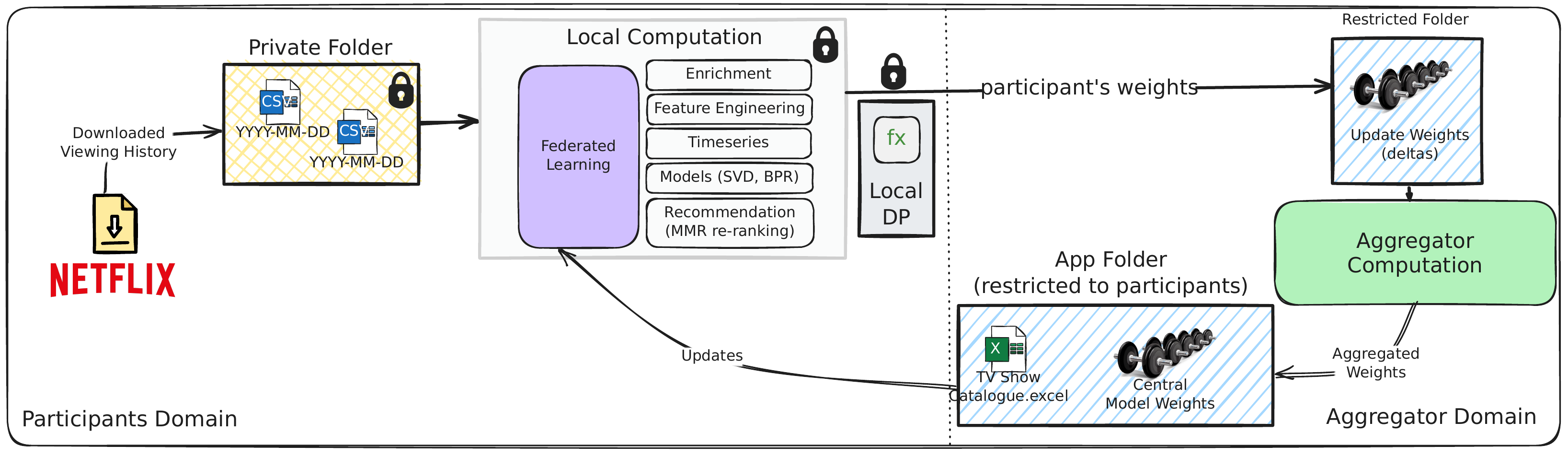}
  \caption{The workflow of FedFlex.}
  \label{fig:fedflex-diagram}
\end{figure}

FedFlex, our application, is a federated recommender for Netflix series. 
Figure~\ref{fig:fedflex-diagram} depicts the flow of data throughout the application, with private data (that is, not accessible by anyone but the user) marked with a lock icon. 
The system consists of two types of actors: live participants and a virtual aggregator. 
The participants represent the users of the system, and the aggregator is the orchestrator of the global model, which is trained on results from the participants' local training.

Understanding the interaction between the aggregator and participants is crucial. The aggregator continuously monitors all shared datasites on the SyftBox system and grants participants access to retrieve the global model once their local application is deployed. With these permissions, participants contribute to model training as described in Section~\ref{sec:finetuning}.

Participants first load their Netflix viewing history into a private local folder. Once the application is running, it converts this history into a vector of ratings for each show, using viewing frequency as a proxy for enjoyment, for example, watching three episodes of a series in a week corresponds to a five-star rating. While this method cannot capture nuances, such as missed episodes due to other commitments, it provides an approximation of explicit feedback from implicit click data. Unwatched items are treated as negatively rated for pairwise comparison models (cf. Section~\ref{sec:modelchoicesbpr}).

Once a vector representing the participant's viewing history is created, it is processed using our recommendation algorithms. Differential privacy is applied to the local training results by adding noise to the model updates shared with the aggregator, preventing inference of user data. The aggregator updates the global model with these shared gradients. Finally, participants generate recommendations locally by combining the updated global model with their own ratings.

\subsection{Data Scraping and Model Initialization}
For model initialization, FedFlex uses static data from whats-on-netflix.com and IMDB collected in December 2024, when the project began. 
The data, scraped using a custom Python script, includes metadata for Netflix series such as descriptions, cast, ratings, and release year. The recommendation model is initialized using the IMDB data: ratings are normalized and used to initialize item vectors, with a small amount of random noise added.

\subsubsection{Recommendation Process}
\label{sec:finetuning}
In this section, we describe each step of the recommendation process and define how user data is processed.

\paragraph{Our SVD-based fine-tuning approach}
FedFlex's fine-tuning approach uses collaborative filtering based on Singular Value Decomposition (SVD). This method leverages explicit feedback to update user and item vectors for individual items. Formally, the local model fine-tuning process proceeds as follows:

Let global item vector \( \boldsymbol{\mathcal{V}} \) be the matrix of item embeddings where \( \boldsymbol{\mathcal{V}}_i \) denotes the embedding of item \( i \), and user vector \( \boldsymbol{\mu}_u \) be the user preferences vector for user \( u \). Given a user rating set \( \boldsymbol{\mathcal{D}}_u = \{(u, i, r_{ui})\} \), where \( r_{ui} \) is the observed rating of item \( i \) by user \( u \), we iteratively update the embeddings for a fixed number of iterations. At each step, we compute the predicted rating \(\boldsymbol{\hat{r}}_{ui} \) for each individual item:
$
\hat{r}_{ui} = \mu_u^\top \mathcal{V}_i$.
The prediction error is calculated as:
$e_{ui} = r_{ui} - \hat{r}_{ui}$.
We then compute the gradients for both the user and item vector, with \( \boldsymbol{\lambda} \) as a regularization parameter:
\[
\nabla_{\mu_u} = e_{ui} \mathcal{V}_i - \lambda \mu_u,
\quad
\nabla_{\mathcal{V}_i} = e_{ui} \mu_u - \lambda \mathcal{V}_i
\]
Finally, we perform the parameter updates using learning rate \( \boldsymbol{\alpha} > 0 \):
\[
\mu_u \leftarrow \mu_u + \alpha \nabla_{\mu_u},
\quad
\mathcal{V}_i \leftarrow \mathcal{V}_i + \alpha \nabla_{\mathcal{V}_i}
\]
This process is repeated for all samples in the training set over a fixed number of iterations. After fine-tuning, the resulting model updates (deltas) are computed, and differential privacy is applied as described in Section~\ref{sec:DP} before sharing them with the aggregator.

\paragraph{Our BPR-based fine-tuning approach}

To provide a more varied baseline, we also implemented a FedBPR\footnote{https://github.com/sisinflab/FedBPR}-inspired algorithm. Bayesian Personalized Ranking (BPR)~\cite{BPR} is a matrix factorization method that uses implicit feedback, comparing positive (clicked) and negative (non-clicked) items. We chose BPR because, like SVD, it is based on matrix factorization and can be integrated without changing input or output dimensions. 
BPR also provides a useful contrast to SVD due to its pairwise update approach.
Formally, we define the process as follows:

Let the global item vector be \( \boldsymbol{\mathcal{V}} \) and the user vector \( \boldsymbol{\mu}_u \). Using the user rating set \( \boldsymbol{\mathcal{D}}_u = \{(u, i, r_{ui})\} \), we split the Netflix series into positives and negatives. The positive set, \( \boldsymbol{\mathcal{I}}_u^+ \), contains series the user has watched, while the negative set, \( \boldsymbol{\mathcal{I}}_u^- \), contains all unwatched series. This split is defined as:
$\mathcal{I}_u^+ = \{ i \mid (u, i, r_{ui}) \in \mathcal{D}_u \}$
Let \( \boldsymbol{\mathcal{I}} = \{0, \dots, n-1\} \) be the full item set, and define the negative item candidates as:
$\mathcal{I}_u^- = \mathcal{I} \setminus \mathcal{I}_u^+$
Then, over a fixed number of iterations, for every positive item \( i \in \boldsymbol{\mathcal{I}}_u^+ \), we sample an item \( j \in \boldsymbol{\mathcal{I}}_u^- \) at random and compute the pairwise predicted rating difference \( \boldsymbol{x}_{uij} \) using the same rating prediction calculation:
\[
x_{uij} = \mu_u^\top \mathcal{V}_i - \mu_u^\top \mathcal{V}_j
\]
We compute the gradient of the pairwise loss using the following sigmoid function, as applied in FedBPR:
\[
\sigma(-x_{uij}) = \frac{1}{1 + e^{x_{uij}}}
\]
We use three separate regularization parameters for the user, positive item, and negative item, denoted by \( \boldsymbol{\lambda}_u \), \( \boldsymbol{\lambda}_i \), and \( \boldsymbol{\lambda}_j \), respectively.
Details are provided in Section~\ref{sec:modelchoicesbpr}. Since implicit data provide less certainty about true negatives, different regularization parameters are required. These parameters are then used to compute the gradients for the user vector and both item vectors via the sigmoid function:
\[
\nabla_{\mu_u} = \sigma(-x_{uij})(\mathcal{V}_i - \mathcal{V}_j) - \lambda_u \mu_u,
\]
\[
\nabla_{\mathcal{V}_i} = \sigma(-x_{uij}) \mu_u - \lambda_i^+ \mathcal{V}_i,
\]
\[
\nabla_{\mathcal{V}_j} = -\sigma(-x_{uij}) \mu_u - \lambda_j^- \mathcal{V}_j
\]
Finally, using learning rate \( \boldsymbol{\alpha} > 0 \), we perform the parameter updates:
\[
\boldsymbol{\mu}_u \leftarrow \boldsymbol{\mu}_u + \alpha \nabla_{\boldsymbol{\mu}_u}, \quad
\boldsymbol{\mathcal{V}}_i \leftarrow \boldsymbol{\mathcal{V}}_i + \alpha \nabla_{\boldsymbol{\mathcal{V}}_i}, \quad
\boldsymbol{\mathcal{V}}_j \leftarrow \boldsymbol{\mathcal{V}}_j + \alpha \nabla_{\boldsymbol{\mathcal{V}}_j}
\]

This process is repeated over a fixed number of iterations, after which the deltas are calculated and differential privacy is applied before sharing the model updates with the aggregator.

\paragraph{Model Aggregation}
Once a participant has performed local fine-tuning and shared the obscured deltas, the aggregator updates the global model. 
Note that this occurs only once per participant, who is then marked as complete to prevent retraining. In a live user study with a small sample size, retraining on the same participant could cause severe overfitting.

For model aggregation, updates are first clipped if necessary to limit the impact of any single update. The aggregated update is then calculated by averaging all updates. 
Differential privacy can be applied as a second layer of protection, controlled by \( \boldsymbol{\epsilon} \), after which the aggregated deltas are added to the global model.

\paragraph{Local Recommendations} are generated by computing all predicted user ratings, as in the SVD and BPR processes, and selecting the top-5 items. To avoid redundancy, previously seen items are filtered out. FedFlex also produces a diverse recommendation set by reranking the predicted ratings.

\subsubsection{Diversity Implementation}
\label{sec:mmr}

This section addresses \textbf{RQ2}, which asks how to promote diversity in a federated recommender system. 
To do so, we use a reranking approach, which is easy to integrate into an existing system.
We apply Maximal Marginal Relevance (MMR)~\cite{MMR} as a straightforward solution.
Formally, we define the MMR reranking process as follows: Let \( \boldsymbol{\mathcal{C}} = \{ (t_i, j_i, \hat{r}_i) \}_{i=1}^{m} \) be a set of \( \boldsymbol{m} \) candidate items, where \( \boldsymbol{t}_i \) is the item's title, \( \boldsymbol{j}_i \) is the item ID, and \( \boldsymbol{\hat{r}}_i \) is the predicted rating. First, we create an embedding \( \boldsymbol{e}_i \) for each item \(i\) to be able to measure similarity (cf. Section~\ref{sec:modelchoicesmmr}).
We normalize the predicted ratings for consistency, with \( \boldsymbol{\tilde{r}_i} \) being the normalized rating for item \( \boldsymbol{i} \):
\[
\tilde{r}_i = \frac{\hat{r}_i - \min_{j \in \mathcal{C}} \hat{r}_j}{\max_{j \in \mathcal{C}} \hat{r}_j - \min_{j \in \mathcal{C}} \hat{r}_j}
\]
To select \( \boldsymbol{N} \) items for diverse recommendations, let \( \boldsymbol{\mathcal{S}} \subseteq \{1, \dots, m\} \) denote the set of already selected item indexes, initially empty. Items are added until \( \boldsymbol{\mathcal{S}} \) contains \( \boldsymbol{N} \) recommendations by choosing the index \( \boldsymbol{i}^* \) that maximizes the MMR score:
\[
i^* = \arg\max_{i \in \mathcal{C} \setminus \mathcal{S}} \left[ \lambda \cdot \tilde{r}_i - (1 - \lambda) \cdot \max_{j \in \mathcal{S}} \cos(\mathbf{e}_i, \mathbf{e}_j) \right]
\]
\( \boldsymbol{\lambda} \in [0, 1] \) controls the trade-off between predicted relevance and dissimilarity. When \( \boldsymbol{\mathcal{S}} = \emptyset \), the diversity penalty is zero, so the first recommended item is simply the one with the highest predicted rating. The formula balances relevance and diversity according to \( \boldsymbol{\lambda} \), comparing each candidate item to the already selected set \( \boldsymbol{\mathcal{S}} \) using cosine similarity and taking the highest similarity as its diversity penalty. Once \( \boldsymbol{N} \) items are selected, the recommendations are presented in descending order of MMR scores.

\subsection{Differential Privacy}
\label{sec:DP}
Differential privacy (DP) is a privacy-enhancing technique to prevent the aggregator server from inferring user data from the shared model updates~\cite{Dwork2008Differential}. 
If model updates were shared directly, a malicious actor could infer the input data. 
DP mitigates this by adding controlled random noise to the updates shared by the user. In FedFlex, DP can also be optionally applied at the aggregator as an additional layer of protection.
Global and local DP differ in where the noise is applied: in global DP, a trusted server adds noise to aggregated results, while in local DP, users add noise before transmitting data. 
We chose local DP because SyftBox applications are open-source, allowing users to verify that their private data is properly privatized and enhancing trust in the system's privacy guarantees.
FedFlex supports Gaussian~\cite{gaussiandp} and Laplace~\cite{laplacedp} noise. 
While other types of DP exist, we chose these distributions because they are widely used and can easily integrate into our implementation to add calibrated randomness to model updates.

\subsection{Web Application for User Study}
\begin{figure}[ht]
  \centering
  \includegraphics[width=0.8\columnwidth]{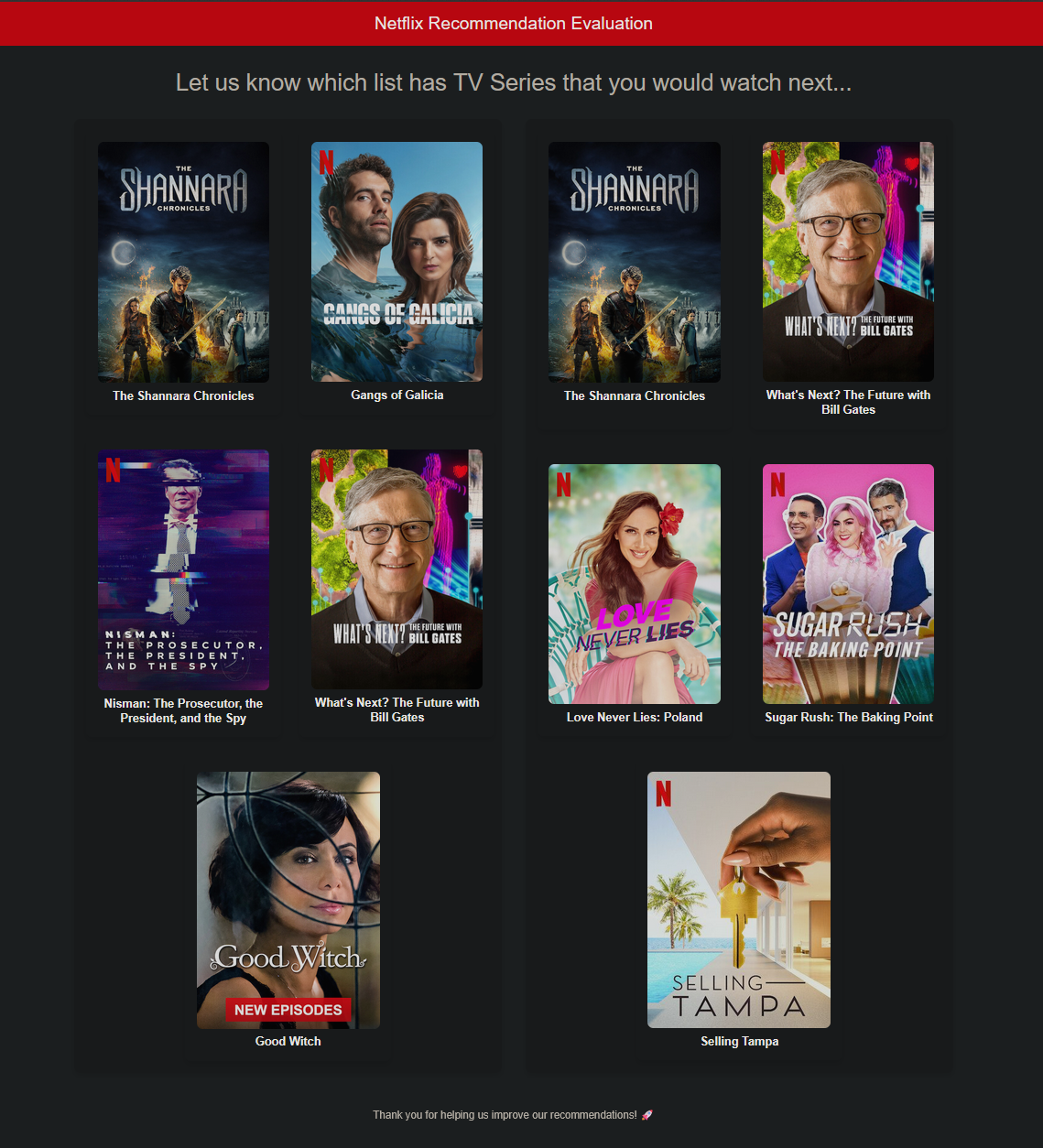}
  \caption{An example of the local web application.}
  \label{fig:webapp}
\end{figure}

To execute the user study, we introduced a web application pictured in Figure~\ref{fig:webapp}. 
This web application runs locally for each participant and displays two columns of five TV series thumbnails. The left column shows the top-5 series by predicted rating, while the right column shows the top-5 after reranking. Displaying both lists simultaneously allows us to collect data on participants' perceived differences between them.

We instruct participants to click on the shows they would be likely to watch. The clicks, timestamps, participant ID, and the two recommendation lists are recorded and shared with the aggregator for this experiment, while participants' viewing histories remain private. 
Participants also complete an exit questionnaire, provided in the supplementary material, to capture insights beyond the experimental metrics. This helps address limitations from the small sample size typical of live user studies and provides an understanding of users' perceptions of the recommendations.

\section{Experimental Setup}
\label{s:evaluation}
In this section, we discuss the model's decisions made for our experiment and how the experiments are performed and what we used to perform them. 

\subsection{Model Choices}
\label{sec:modelchoices}

\paragraph{Fine-Tuning}
\label{sec:modelchoicesbpr}
SVD and BPR have three main parameters. The first is the learning rate \( \boldsymbol{\alpha} \), which controls the impact of each gradient update. It ranges from 0 (no updates) to 1 (full updates). For this experiment, we used a standard value of \( \boldsymbol{\alpha} = 0.01 \), which is low but suitable given our small-scale data.

Secondly, we have the number of iterations for model training. 
We chose 10 iterations as a default. 
Finally, we have the regularization parameter \( \boldsymbol{\lambda} \). For SVD, we use a \( \boldsymbol{\lambda} \) = 0.1.
For BPR, regularization is split between a regularization parameter for user vector updates, positive item updates, and negative item updates as \( \boldsymbol{\lambda}_u \), \( \boldsymbol{\lambda}_i \), and \( \boldsymbol{\lambda}_j \), respectively. 
Here, we set the parameters to have the same values as the FedBPR implementation~\cite{usercontrolfedtopn}.  


\paragraph{Promoting diversity using maximal marginal relevance}
\label{sec:modelchoicesmmr}
MMR must be configured for each use case, including a method to measure similarity between items and a trade-off between relevance and diversity. We use title-based embeddings generated with the \textit{sentence-transformers} module (\textit{all-MiniLM-L6-v2}) and compute similarity using cosine similarity between the embeddings.
The parameter \( \boldsymbol{\lambda} \), which controls the trade-off between diversity and relevance, is set to a low value of 0.3. Lambda was tuned through offline tests before the live study. In this experiment, diversity is emphasized: predicted ratings contribute 30\% to an item's score, while the similarity penalty contributes 70\%.



\paragraph{Differential privacy}

Our differential privacy implementation uses \( \boldsymbol{\epsilon} \) as the privacy budget, along with a sensitivity parameter and a noise type. Smaller values of \( \boldsymbol{\epsilon} \) correspond to more noise added to the model updates.
We used a standard privacy budget of \( \boldsymbol{\epsilon} = 1 \) and a sensitivity of 0.36 for this experiment. FedFlex supports both Gaussian and Laplace noise, and we used Gaussian by default. Future work will explore the optimal choice of \( \boldsymbol{\epsilon} \).

\subsection{User Study}

For this experiment, we invited participants through family, friends, and acquaintances using various channels, such as social networks (e.g., Slack) and individual requests, without offering any incentive. Instructions were provided via text and video to guide participants in retrieving their Netflix history, installing the FedFlex app, and completing the experiment.

Over a period of two weeks, 13 participants ran the experiment at least once (i.e., sharing their data to train the global model), with 10 completing the full experiment by sharing their preferred clicks with us. 
The user study consists of the participants running two versions of FedFlex, one with the SVD-based algorithm and the other with the BPR-based algorithm. 
We split the apps into two separate versions to make it easier to train two separate global models, which are publicly available as part of our
contributions. 
The participants were then asked to run the application several times over the span of two weeks and input their clicks for the series they would like to see.
After two weeks of the experiment, we asked the users to fill out a questionnaire. 
The questions focused on five categories, namely, (1) the overall user satisfaction, (2) perceived diversity between the regular and reranked lists, (3) perception of the content of the recommendations, (4) federated learning and privacy, and (5) Final comments on the study.
This questionnaire is mainly to gain additional insights into the subjective metrics of how each user felt about their experience with the recommendations, what they noticed about it, and to hint at what we could improve for future research.

\subsection{Metrics}

To evaluate our system, we need a set of metrics for both accuracy and diversity measurements (\textbf{RQ3}). 
For our \textbf{accuracy} metrics, we choose Click-Through Rate (CTR) and Normalized Discounted Cumulative Gain (nDCG@K).
CTR measures how many of the recommended items were clicked:

\[
CTR = \frac{\text{Number of Clicks}}{\text{Number of Recommended Items}}
\]


nDCG@K evaluates positional relevance, assuming earlier clicks indicate stronger interest:
\[ 
\mathrm{nDCG}@k = \frac{\mathrm{DCG}@k}{\mathrm{IDCG}@k}, 
\]
\[
\mathrm{DCG}@k = \sum_{i=1}^{k} \frac{2^{\mathrm{rel}_i}-1}{\log_2(i+1)}, \quad 
\mathrm{IDCG}@k = \sum_{i=1}^{k} \frac{2^{\mathrm{rel}^*_i}-1}{\log_2(i+1)}
\]
where \(\mathrm{rel}_i\) and \(\mathrm{rel}^*_i\) are the relevance scores at position \(i\) in the predicted and ideal rankings, respectively, and \(k\) is the number of recommended items.



For our \textbf{diversity}, we opted for Intra-List Diversity, Coverage Ratio, and KL divergence. 




\textit{Intra-List Diversity (ILD)} measures how diverse a recommendation list is by capturing the average dissimilarity between recommended items. In this context, dissimilarity is based on the genres of the items: the more different the genres, the higher the ILD. 
A higher ILD indicates that the recommendations cover a wider variety of genres.
\textit{Coverage Ratio (CR)} describes the ratio between the number of unique items clicked and the number of unique items recommended:
\textit{KL divergence (KL)} measures how different the distribution of genres in the clicked recommendations is from the overall reference distribution of genres in the catalog. 
A lower value indicates that the recommendations cover genres more proportionally to the catalog (higher diversity), while a higher value suggests over-representation of certain genres (lower diversity). 

\section{Experimental Results}
In this section, we present our experimental results, focusing on recommendation accuracy, diversity, and the exit questionnaire.
We also discuss the nuance behind these results and how they could be improved. 

\subsection{Recommendations Accuracy}

\begin{figure}[!ht]
\centering

    \includegraphics[width=\columnwidth]{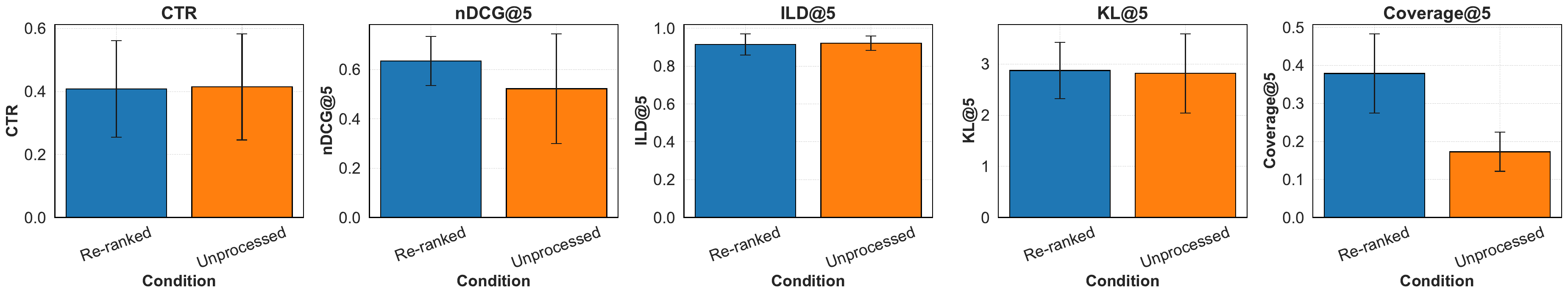}
    
    \includegraphics[width=\columnwidth]{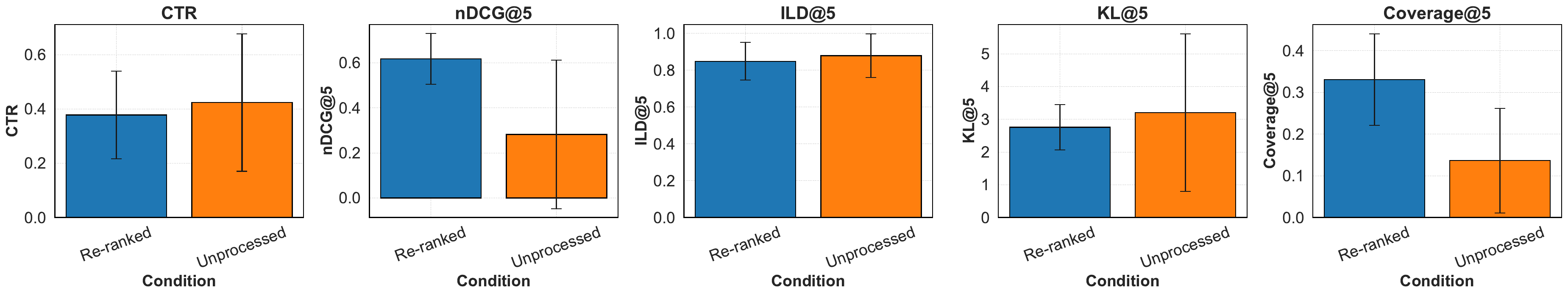}
    \caption{Recommendation accuracy and diversity with mean and confidence interval for both the SVD (\texttt{top}) and BPR (\texttt{Bottom}) experiments.}
    \label{fig:BPR-SVD_metrics}
\end{figure}

As part of our experiments, we first aimed to determine whether participants would find our recommended TV series interesting and click on them. 
We conducted statistical analyses using independent two-tailed t-tests, with Cohen's d as the effect size and 95\% confidence intervals for the mean differences. 
In Table~\ref{tab:performance} and Figure~\ref{fig:BPR-SVD_metrics}, we present our recommendation results measured in terms of CTR and nDCG@5. 
\begin{table}[t]
\centering
\scriptsize
\setlength{\tabcolsep}{4pt} 
\caption{Recommendation performance: accuracy (CTR, nDCG) and diversity (KL, ILD, CR). 
The cutoff is 5. Arrows indicate increase/decrease compared to baseline. 
Significance: \textsuperscript{*} $p<0.05$, \textsuperscript{†} $p<0.01$.}
\begin{tabular}{l c c c c c}
\toprule
\multirow{2}{*}{\textbf{Model}} & \multicolumn{2}{c}{\textbf{Accuracy}} & \multicolumn{3}{c}{\textbf{Diversity}} \\
\cmidrule(lr){2-3} \cmidrule(lr){4-6}
 & \textbf{CTR} & \textbf{nDCG} & \textbf{KL} & \textbf{ILD} & \textbf{CR} \\
\midrule
SVD        & \bfseries 0.415 & 0.522 & \bfseries 2.819 & \bfseries 0.921 & 0.173 \\
SVD+MMR    & 0.408 {\small ↓} & \bfseries 0.634 {\small ↑} & 2.873 {\small ↑} & 0.915 {\small ↓} & \bfseries 0.379 {\small ↑\textsuperscript{†}} \\
\midrule
BPR        & \bfseries 0.424 & 0.282 & 3.204 & 0.848 & 0.137 \\
BPR+MMR    & 0.378 {\small ↓} & \bfseries 0.618 {\small ↑\textsuperscript{*}} & \bfseries 2.759 {\small ↓} & \bfseries 0.879 {\small ↑} & \bfseries 0.331 {\small ↑\textsuperscript{†}} \\
\bottomrule
\end{tabular}
\label{tab:performance}
\end{table}

Overall, the effects of re-ranking on accuracy were mixed. 
For BPR, applying MMR significantly improved ranking quality (nDCG@5: M\_diff = 0.358, t = -2.73, \textit{p = 0.034}, d = 1.03), suggesting that re-ranking can improve accuracy for this model. 
In contrast, SVD outperformed its re-ranked counterpart (SVD+MMR), and CTR tended to decrease slightly after re-ranking.
Taken together, these results indicate that while BPR+MMR benefits from re-ranking, accuracy outcomes across models are inconsistent and largely non-significant. 
This suggests that re-ranking does not lead to a systematic loss in accuracy, but its improvements may depend on the underlying model (\textbf{RQ3}).
The variability in the results is likely influenced by two factors: the relatively small dataset and the encoding of interactions as implicit in BPR (e.g., clicks) versus explicit in SVD (e.g., ratings).
Future work with larger samples and dynamic user modeling is expected to yield more robust conclusions.

\subsection{Recommendations Diversity}
Second, we measure diversity by applying MMR, which promotes variety by re-ranking items at the top of the recommendation list. 
The goal is to examine whether participants engage with these more diverse recommendations. 
As shown in Table~\ref{tab:performance} and Figure~\ref{fig:BPR-SVD_metrics}, the results are mixed. 
For SVD vs. SVD+MMR, intra-list diversity (ILD) remained unchanged (M\_diff = 0.007, t = 0.61, p = 0.56, d = 0.19), while KL divergence slightly worsened (M\_diff = +0.256, t = 1.63, p = 0.18, d = 0.73). 
However, coverage (CR) improved significantly (M\_diff = 0.206, t = -4.20, $p < 0.01$, d = -1.33), suggesting that MMR broadened the pool of recommended items. 
For BPR vs. BPR+MMR, re-ranking also yielded a significant coverage gain (M\_diff = 0.175, t = -5.97, $p = 0.001$, d = 2.25), while ILD remained stable (M\_diff = 0.010, t = 1.00, p = 0.36, d = 0.38). 
Notably, KL divergence decreased (M\_diff = -0.445, t = -1.00, p = 0.42, d = -0.58), indicating a trend toward more diverse recommendations.
These differences highlight that coverage, ILD, and KL capture distinct aspects of diversity and do not necessarily move in the same direction.


    
\begin{figure*}[]
    \centering
    \includegraphics[width=\textwidth]{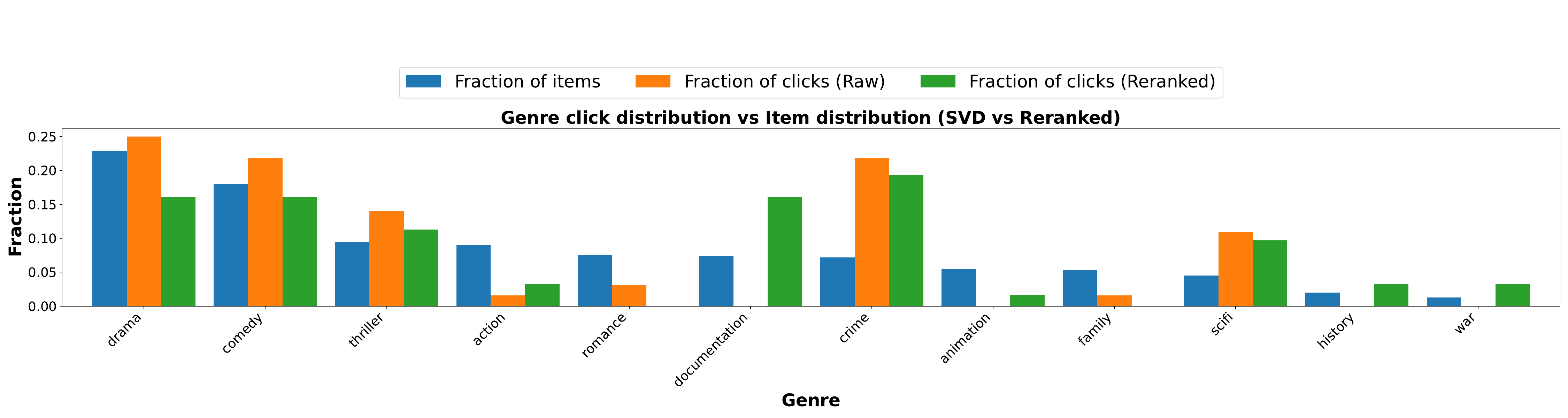}

    \vspace{-0.5cm} 

    \includegraphics[width=\textwidth]{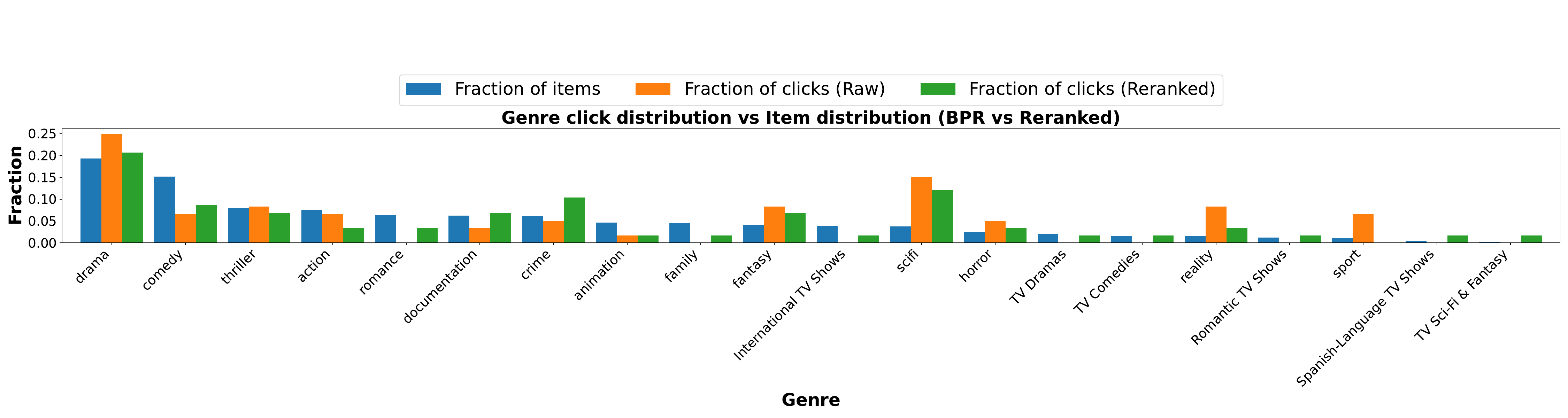}
    
    \caption{Distribution of clicked genres between raw and reranked recommendations for both the SVD and BPR.}
    \label{fig:genre_click_dist}
\end{figure*}

Figure~\ref{fig:genre_click_dist} shows the difference in click distribution between the raw and reranked recommendation lists for each experiment. 
As mentioned before, our analysis shows that we succeed in generating more diverse recommendations at the individual item level. 
Combining our experiment data with the Netflix series data, we further analyze whether this is also the case for genres. 
We can see that the reranked recommendations consistently generate clicks for new genres for the users, whilst still adhering to strong preferences (i.e., if a user has many high-scoring items in the drama genre, the reranked recommendation clicks will still mostly contain drama). 
Note that especially reranked BPR introduces many new genres compared to SVD, which only does so in a very limited manner. This is mainly due to the difference between SVD and BPR. During experimentation, we noticed that due to the nature of BPR, being based on pairwise comparison and applying a sigmoid gradient, predicted ratings were much more bounded compared to SVD, which would often have very high individual ratings. 
This also means that SVD would occasionally rate shows for users very high and overwhelm the reranker to an extent, which explains why, in the case of SVD, the reranker sticks closer to the original recommended genres.



\subsection{Exit Questionnaire}
The questionnaire revealed that most users were unsure whether the recommendations were personalized, likely due to the small sample size and the cold start problem.
The majority expressed no clear preference between the two lists, suggesting that increased diversity did not substantially reduce satisfaction.
One user, for instance, rated their likelihood of watching a recommended series as 4 out of 5. They noted differences between the lists, perceiving the reranked list as more aligned with their interests. They also encountered novel series, though not always positively. While they did not feel their data was private, they were comfortable with participation and open to joining similar studies in the future.

This example points to several directions for improvement. Pre-training the model could strengthen personalization and satisfaction. While diversification was valuable for this user, preferences may vary, pointing to the need for dynamic parameters. Finally, the mixed reception of novelty suggests exploring alternative similarity measures (e.g., incorporating genre) to refine how diversity is promoted.

\section{Discussion}
\label{s:discussion}
We view the results as an indication that our initial experiments were successful and hold potential at a larger scale.
That said, the metrics should be interpreted with caution, as the cold start problem and small sample size limit personalization compared to large-scale studies.
Still, our findings suggest that the approach is viable for (1) deployment in a federated learning live study and (2) achieving more diverse recommendations, with further research needed to confirm scalability.
It remains important to contextualize the results and reflect on how they emerged (\textbf{RQ3}).
To achieve these results and commit to further study, it is important to discuss the nuances of our approach.
A key trade-off in live studies lies between re-training on user data and avoiding overfitting. We chose not to re-train the model during the two-week study, as the small sample size risked overfitting to users who submitted new viewing histories more frequently. Instead, each user was trained once, which reduced adaptivity but ensured comparability across participants.

Ideally, this trade-off could be handled dynamically: the system would re-train when substantial novel data are added, balancing personalization gains against overfitting risks. In federated learning, however, this decision must happen locally on the client side, where malicious users could exploit retraining triggers. The problem is further compounded when differential privacy is applied, as each update consumes part of the global privacy budget $\epsilon$. Frequent retraining would not only increase the risk of overfitting but also accelerate privacy loss, leaving less budget for future updates and potentially limiting the system's long-term viability.

It is also important for user experience that recommendations are not redundant, i.e., show the user has already seen. In our setup, we filtered out series present in the user-submitted data. However, if a user had watched a new series without updating their data, the system could still recommend already-seen content. While this was a limitation of our experimental setup, a real-world system should automatically update and process the latest viewing history to avoid such cases.

We believe FedFlex successfully addresses ethical concerns in a context where user data is often treated as a commodity rather than a user's property. By giving users control over what data are processed and keeping details fully private, FedFlex demonstrates that federated technologies can ensure privacy and foster diversity in recommendations. Our results suggest that FedFlex can contribute to the future of series recommendations by delivering accurate and varied results without compromising data privacy. Nevertheless, significant improvements are needed before FedFlex becomes a viable alternative in real-world scenarios. 

\section{Conclusion and Future Work}
\label{futurework}

In this work, we propose \textbf{FedFlex}, to the best of our knowledge, the first live federated recommender study for Netflix-style TV series. FedFlex uses BPR and SVD for personalized fine-tuning and applies Maximal Marginal Relevance (MMR) to re-rank items and improve diversity. Our results demonstrate that we can deliver accurate recommendations while keeping user click data local. MMR also contributes to the promotion of more diverse recommendations.


Even though our study used a small sample size of participants, it sets the promising direction for a larger-scale, longitudinal deployment could offer deeper insights. Additionally, we believe that click-based evaluation would benefit from pre-trained models with personalized parameters (e.g., user-specific MMR $\lambda$). Future work could also incorporate richer data, such as explicit ratings and enhanced similarity features using cast and genre metadata, to improve personalization and diversity further.

\appendix
\section{Checklist}

\begin{enumerate}
    \item For all models and algorithms presented, check if you include:
    \begin{itemize}
        \item A clear description of the mathematical setting, assumptions, algorithm, and/or model. [Yes]
        \item An analysis of the properties and complexity (time, space, sample size) of any algorithm. [Not Applicable]
        \item (Optional) Anonymized source code, with specification of all dependencies, including external libraries. [Yes]
    \end{itemize}

    \item For any theoretical claim, check if you include:
    \begin{itemize}
        \item Statements of the full set of assumptions of all theoretical results. [Not Applicable]
        \item Complete proofs of all theoretical results. [Not Applicable]
        \item Clear explanations of any assumptions. [Yes]
    \end{itemize}

    \item For all figures and tables that present empirical results, check if you include:
    \begin{itemize}
        \item The code, data, and instructions needed to reproduce the main experimental results (either in the supplemental material or as a URL). [Yes]
        \item All the training details (e.g., data splits, hyperparameters, how they were chosen). [Yes]
        \item A clear definition of the specific measure or statistics and error bars (e.g., with respect to the random seed after running experiments multiple times). [Yes]
        \item A description of the computing infrastructure used (e.g., type of GPUs, internal cluster, or cloud provider). [Yes]
    \end{itemize}

    \item If you are using existing assets (e.g., code, data, models) or curating/releasing new assets, check if you include:
    \begin{itemize}
        \item Citations of the creator if your work uses existing assets. [Yes]
        \item The license information of the assets, if applicable. [Yes]
        \item New assets either in the supplemental material or as a URL, if applicable. [Yes]
        \item Information about consent from data providers/curators. [No] None of the providers of the datasets we used provides information about consent.
        \item Discussion of sensible content if applicable, e.g., personally identifiable information or offensive content. [Not Applicable]
    \end{itemize}

    \item If you used crowdsourcing or conducted research with human subjects, check if you include:
    \begin{itemize}
        \item The full text of instructions given to participants and screenshots. [Not Applicable]
        \item Descriptions of potential participant risks, with links to Institutional Review Board (IRB) approvals if applicable. [Not Applicable]
        \item The estimated hourly wage paid to participants and the total amount spent on participant compensation. [Not Applicable]
    \end{itemize}
\end{enumerate}

\section{Exit Questionnaire}
\begin{enumerate}

    \item \textbf{Overall satisfaction}
    \begin{enumerate}
        \item How satisfied were you with the series recommendations overall?\\
        \emph{Likert scale}
        \item How likely are you to watch series based on these recommendations?\\
        \emph{Likert scale}
        \item Did you feel the recommendations were personalized to your taste?\\
        \emph{Yes / No / Not sure}
    \end{enumerate}


    \item \textbf{Perceived differences between algorithms}
    \begin{enumerate}
        \item Did you notice any differences between the two types of recommendation lists you saw (repeatedly)?\\
        \emph{Yes / No / I don't remember}
        \item (If Yes) What kind of differences did you notice?\\
        \emph{Open text}
        \item Which list did you prefer overall?\\
        \emph{List A / List B / No preference}
    \end{enumerate}


    \item \textbf{Content perception}
    \begin{enumerate}
        \item Did the recommendations include series you hadn't heard of before?\\
        \emph{Yes / No}
        \item Did any recommendation surprise you in a good way?\\
        \emph{Yes / No}
        \item If yes: Please give an example?\\
        \emph{Open text}
    \end{enumerate}


    \item \textbf{Federated learning and privacy}
    \begin{enumerate}
        \item Did you feel your data was kept private during the study?\\
        \emph{Yes / No}
        \item How comfortable were you with participating in a recommendation study using decentralized learning (federated learning)?\\
        \emph{Likert scale}
    \end{enumerate}

    \item \textbf{Final comments}
    \begin{enumerate}
        \item Would you like to participate in a similar study again?\\
        \emph{Yes / No / Maybe}
        \item If you have any further comments or feedback on the study, please leave them here\\
        \emph{Open text}
    \end{enumerate}

\end{enumerate}

\begin{acks}
The authors would like to thank the OpenMined Community for their valuable time in answering our questions. Also, we thank all participants for taking part in our experiments. 
\end{acks}

\bibliographystyle{ACM-Reference-Format}
\bibliography{6_bibliography}










\end{document}